\documentclass[a4paper,fleqn]{cas-dc}

\usepackage[numbers]{natbib}
\usepackage{xcolor}
\usepackage{tabularx}
\usepackage{xspace}
\usepackage{listings}
\usepackage{booktabs}
\usepackage{subfigure}
\usepackage{pythonhighlight}
\usepackage{amsfonts}
\usepackage{amsmath,amssymb}
\usepackage{multirow}
\usepackage{multicol}
\usepackage{float}
\usepackage[ruled,vlined]{algorithm2e}
\usepackage[frozencache,cachedir=.]{minted}

\usepackage{tcolorbox}
\tcbuselibrary{minted,breakable,xparse,skins}

\definecolor{bg}{gray}{0.99}
\DeclareTCBListing{mintedbox}{O{}m!O{}}{%
  breakable=true,
  listing engine=minted,
  listing only,
  minted language=#2,
  minted style=default,
  minted options={%
    linenos,
    gobble=0,
    breaklines=true,
    breakafter=,,
    fontsize=\small,
    numbersep=8pt,
    #1},
  boxsep=0pt,
  left skip=0pt,
  right skip=0pt,
  left=15pt,
  right=0pt,
  top=3pt,
  bottom=3pt,
  arc=5pt,
  leftrule=0pt,
  rightrule=0pt,
  bottomrule=1pt,
  toprule=1pt,
  colback=bg,
  colframe=orange!70,
  enhanced,
  overlay={%
    \begin{tcbclipinterior}
    \fill[orange!20!white] (frame.south west) rectangle ([xshift=10pt]frame.north west);
    \end{tcbclipinterior}},
  #3}

\newcommand{\codeword}[1]{%
\texttt{\textcolor{blue}{#1}}%
}
\newcommand{\tool}{CRaDLe\xspace}

\def\tsc#1{\csdef{#1}{\textsc{\lowercase{#1}}\xspace}}
\tsc{WGM}
\tsc{QE}
\tsc{EP}
\tsc{PMS}
\tsc{BEC}
\tsc{DE}
\begin{document}
\let\WriteBookmarks\relax
\def\floatpagepagefraction{1}
\def\textpagefraction{.001}
\shorttitle{\tool}
\shortauthors{Wenchao~GU~et~al.}

\title [mode = title]{\tool: Deep Code Retrieval Based on Semantic Dependency Learning}

\author[1]{Wenchao~Gu}[orcid=0000-0003-3503-8845]
\ead{wcgu@cse.cuhk.edu.hk}

\address[1]{The Department of Computer Science and Engineering, The Chinese University of Hong Kong, Hong Kong, China}

\author[2]{Zongjie~Li}
\ead{lizongjie@stu.hit.edu.cn}

\author[2]{Cuiyun~Gao}
\cormark[1]
\ead{gaocuiyun@hit.edu.cn}

\address[2]{The School of Computer Science and Technology, Harbin Institute of Technology, Shenzhen, China}

\author%
[2]
{Chaozheng~Wang}
\ead{wangchaozheng@stu.hit.edu.cn}

\author%
[3]
{Hongyu~Zhang}
\ead{hongyu.zhang@newcastle.edu.au}

\author%
[2]
{Zenglin~Xu}
\ead{xuzenglin@hit.edu.cn}

\author%
[1]
{Michael~R.~Lyu}
\ead{lyu@cse.cuhk.edu.hk}

\address[3]{The University of Newcastle, Australia}

\cortext[cor1]{Corresponding author}
% \cortext[cor2]{Principal corresponding author}

\begin{abstract}
Code retrieval is a common practice for programmers to reuse existing code snippets in the open-source repositories. Given a user query (i.e., a natural language description), code retrieval aims at searching the most relevant ones from a set of code snippets. The main challenge of effective code retrieval lies in mitigating the semantic gap between natural language descriptions and code snippets. With the ever-increasing amount of available open-source code, recent studies resort to neural networks to learn the semantic matching relationships between the two sources. The statement-level dependency information, which highlights the dependency relations among the program statements during the execution, reflects the structural importance of one statement in the code, which is favorable for accurately capturing the code semantics but has never been explored for the code retrieval task.
In this paper, we propose \tool, a novel approach for \textbf{C}ode \textbf{R}etrieval based on statement-level sem\textbf{a}ntic \textbf{D}ependency \textbf{Le}arning. Specifically, \tool distills code representations through fusing both the dependency and semantic information at the statement level, and then learns a unified vector representation for each code and description pair for modeling the matching relationship. Comprehensive experiments and analysis on real-world datasets show that the proposed approach can accurately retrieve code snippets for a given query and significantly outperform the state-of-the-art approaches on the task.
\end{abstract}

% \begin{highlights}
% \item Research highlights item 1
% \item Research highlights item 2
% \item Research highlights item 3
% \end{highlights}


\maketitle
\section{Introduction}
Implementing projects from scratch is tedious for programmers. In most cases, they know what they want to do, but do not have the capability to implement all the details. For example, a Python programmer may want to ``\textit{convert date\_string into datetime format}'', but not able to recognize the proper syntax \codeword{datetime.strptime(date\_string, format)} for the realization. To mitigate the impasse, it is common for programmers to search the web in natural language (NL), find relevant code snippets, and modify them into the desired form~\citep{DBLP:conf/chi/BrandtGLDK09}. Many code retrieval approaches~\citep{DBLP:conf/chi/BrandtGLDK09,DBLP:conf/kbse/LvZLWZZ15, DBLP:conf/icse/McMillanGPXF11} have been proposed to improve the recommendation accuracy of the returned code snippets given a natural language description. The main challenge of effective code retrieval is the semantic gap between source code and natural language descriptions since the two sources are heterogeneous and share few common lexical tokens, synonyms and language structures~\citep{DBLP:conf/icse/GuZ018}.

Prior efforts have been conducted for effective code retrieval. The existing research can be divided into two categories according to the involved techniques, i.e., Information Retrieval (IR)-based and Deep Neural Network (DNN)-based. The IR-based techniques rely on token-wise similarities between source code and queries. Since the variable and API definitions in code are generally word combinations or abbreviations in natural language, more semantically-similar tokens in code and queries can indicate more relevancy between them. For example, McMillan et al. propose Portfolio which utilizes keyword matching and PageRank to return a list of functions \citep{DBLP:conf/icse/McMillanGPXF11}. Lv et al. propose CodeHow to combine API matching for code retrieval~\citep{DBLP:conf/kbse/LvZLWZZ15}. With an increasing amount of available source code and flourish development of deep learning techniques, many studies~\citep{DBLP:conf/icse/GuZ018, DBLP:journals/corr/abs-1909-09436} propose to adopt neural network models for jointly encoding tokens of source code and queries into a single and joint vector space, where one encoder is employed for each input (natural or programming) sequence. The objective is to map semantically relevant code and language into vectors that are near to each other in the vector space. 

Considering the highly-structured characteristic of source code, recent research proposes to integrate the structural information of code such as Abstract Syntax Tree (AST) and Control Flow Graph (CFG) for representing code semantics~\citep{DBLP:conf/acl/YinN17,DBLP:conf/kbse/WanSSXZ0Y19,DBLP:conf/icse/ZhangWZ0WL19}, demonstrating the effectiveness of involving structural information for the task. However, the deep nature of the extracted trees in ASTs renders it hard for deep learning models to comprehensively capture the structural information~\citep{DBLP:conf/icse/ZhangWZ0WL19}.
% code semantics 
CFG, which represents all possible execution paths for a program, may contain statement orders which are not contributing to the actual execution result, probably leading to biased code representation learning \citep{DBLP:conf/kbse/WanSSXZ0Y19}. In this paper, we propose to utilize statement-level dependency relations in a code snippet based on Program Dependency Graph (PDG). The PDG is established based on AST but less deeper than AST in the structure and only retains the execution paths that will affect the execution result. The dependency relations are then explicitly integrated with the statement-level semantics to capture the code semantics. Actually, the effectiveness of incorporating dependency relations for code representation learning has proven in tasks such as bug detection~\citep{DBLP:journals/pacmpl/LiWNN19} and code clone detection~\citep{DBLP:conf/sigsoft/HendersonP16}; while no prior work has explored the impact on the code retrieval task so far.

Specifically, we introduce a novel neural network model named \tool, an abbreviation of \textbf{C}ode \textbf{R}etrieval based on sem\textbf{a}ntic \textbf{D}ependency \textbf{Le}arning. \tool couples both structural and semantic information of code at the statement level, where the code structures are extracted based on PDG. Extensive experiments have been conducted to verify the performance of the proposed approach. The evaluation results show that \tool can significantly outperform the state-of-the-art models by at least 36.38\% and 22.34\% on two real datasets respectively, in terms of R@1, one standard metric for validating recommendation performance.

In summary, the main contributions of the paper include:
\begin{itemize}
    \item We propose a novel code retrieval model, \tool, to encode both source code and natural language queries into unified vector representations. \tool is the first code retrieval approach that integrates the dependency and semantics information at the statement level for learning code representations.
    \item We conduct large-scale experimental evaluations on public benchmarks. The results demonstrate the superior performance of \tool over the state-of-the-art and baseline models.
\end{itemize}

The rest of this paper is organized as follows. Section~\ref{sec:tool} introduces an overview of the proposed approach and details the design of the approach. Section~\ref{sec:eval} illustrates the experimental datasets, evaluation metrics, and implementation details. Section~\ref{sec:result} elaborates on the experimental results. Section~\ref{sec:literature} surveys the related work and Section~\ref{sec:conclusion} concludes our work.

\section{The Proposed \tool}\label{sec:tool}
In this section, we elaborate on the overview and detailed design of the proposed approach \tool, including the code encoder, description encoder and the similarity measurement component.

\subsection{Overview}\label{sec:method}

\begin{figure}[ht]
\centering
\includegraphics[width=0.5\textwidth]{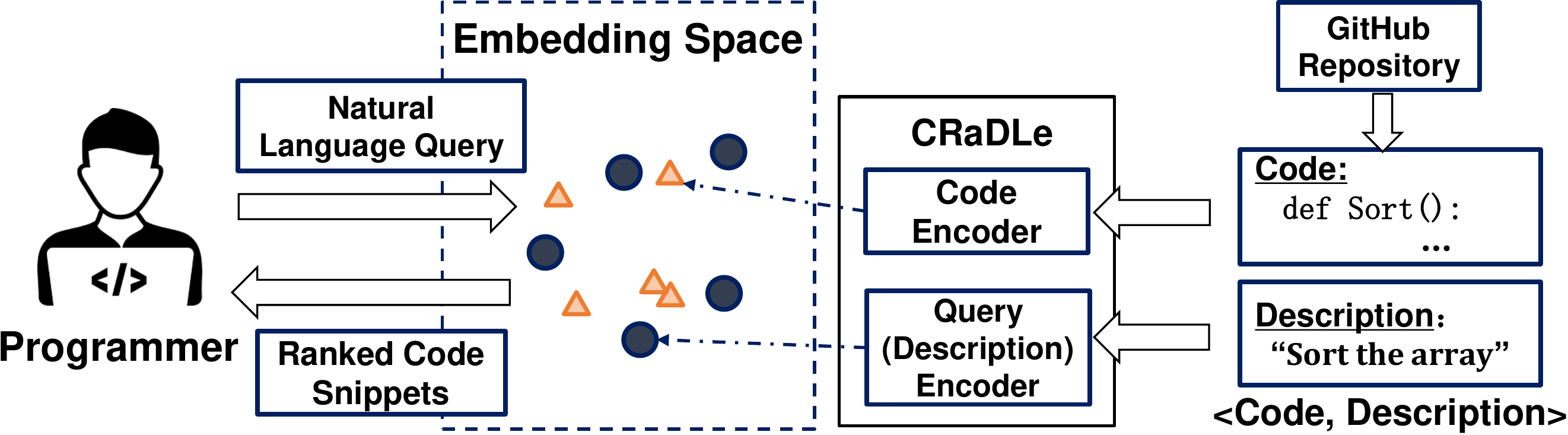}
\caption{Overview of the proposed \tool.}
\label{fig:workflow}
\end{figure}

Figure~\ref{fig:workflow} depicts the overview of the proposed approach, \tool. The implementation includes both offline and online modes. During the offline stage, we first collect datasets containing \textless code, description\textgreater\, pairs. The collected code and descriptions are then preprocessed and separately encoded into vectors by the code encoder and query encoder respectively. Unified representations of code and corresponding descriptions are finally learnt after the offline training process, where semantically similar code and descriptions locate closely to each other in the same embedding space. During the online process, when a new natural language query arrives, the trained model recommends the most related code snippets to the programmer according to the semantic distances between code and the query in the embedding space.

\begin{figure*}[ht]
\centering
\includegraphics[width=0.8\textwidth]{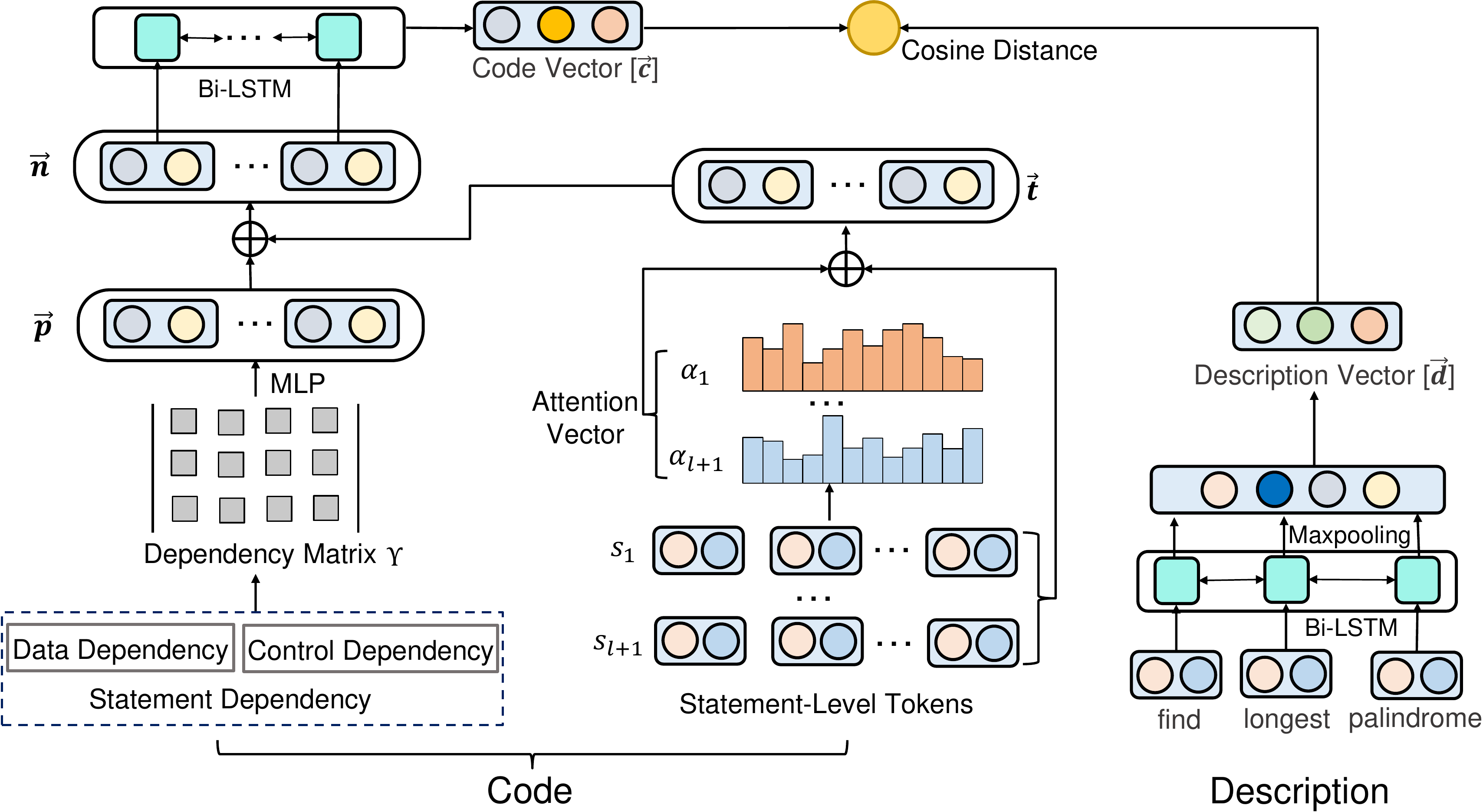}
\caption{Overall framework of the proposed \tool.}
\label{fig:framework}
\end{figure*}

Figure~\ref{fig:framework} illustrates the overall framework of the \tool approach, which details the design of the code encoder and description encoder. The code encoder fuses the statement-level token semantics and distilled dependency information to represent the code semantics. The description encoder also embeds the token sequences in the descriptions to vectors. Finally, similarity matching scores between the code and descriptions are learnt based on their respective vector representations.

\subsection{Code Encoder}
The code encoder aims at embedding code snippets into vector representations. We propose to integrate the statement-level token semantics with the dependency information between statements for accurately capturing the code semantics. We first illustrate the process conducted for the dependency information extraction, and then describe the networks proposed for learning statement-level dependency and semantic representations.

Algorithm~\ref{algo1} shows the procedures for the code encoder. The input of the code encoder includes the token matrix $E$ comprised by a sequence of token embedding vectors $\{\mathbf{e}_{1,1},...,\mathbf{e}_{i,j},...\}$ and dependency matrix $\Upsilon$. First, the dependency embedding layer encodes the dependency matrix $\Upsilon$ into dependency embeddings $P$. The token embedding layer then represents the token matrix $E$ into statement-level representations $T$ comprised by $\{\mathbf{t}_{1},...,\mathbf{t}_{i},...\}$. Finally, the token embeddings are concatenated with the dependency embeddings in statement level, and the newly comprised vectors are fed into the Bi-LSTM layer. The last hidden state vector from the Bi-LSTM layer is treated as the representation vector of the code.

\begin{algorithm}
\SetAlgoLined
\SetKwFunction{FCODEENCODER}{CODEENCODER}
\SetKwProg{Fn}{Function}{:}{}
\SetKwInOut{Input}{input}
\SetKwInOut{Output}{output}
\Input{the token matrix $E$, the matrix of input dependency: $\Upsilon$}
\Output{The representation vector of code: $C$}
 \Fn{\FCODEENCODER{$E$, $\Upsilon$}}{
 $P \leftarrow DependencyEmbedding(\Upsilon)$ \tcp*{corresponding to Equ.~\ref{con:dependencyembedding}}
 $T \leftarrow TokenEmbedding($E$)$\;
 $S \leftarrow StatementAttention(T)$ \tcp*{corresponding to Equ.~\ref{con:attn1} and Equ.~\ref{con:attn2}}
 $C \leftarrow SemanticDependencyEmbedding([S; P])$ \tcp*{corresponding to Equ.~\ref{con:lstm}}
 \Return $C$\;
 }

 \caption{The algorithm of code encoding}
 \label{algo1}
\end{algorithm}

\subsubsection{Dependency Information Extraction}\label{sec:pdg}
We obtain the dependency information between statements by adopting PDGs of the code snippets. PDG explicitly indicates the data dependency and control dependency of a program, where the data dependency can represent the relevant data flow relationships and control dependency exhibits the essential control flow relationships~\citep{DBLP:journals/toplas/FerranteOW87}. Since there exists no mature tool for extracting PDG of one code snippet in interpreted languages such as Python, we propose to establish the PDG based on the AST of a code snippet.

\begin{figure*}
	\centering
	\begin{tabular}{ccc}
	\includegraphics[width=0.3 \textwidth]{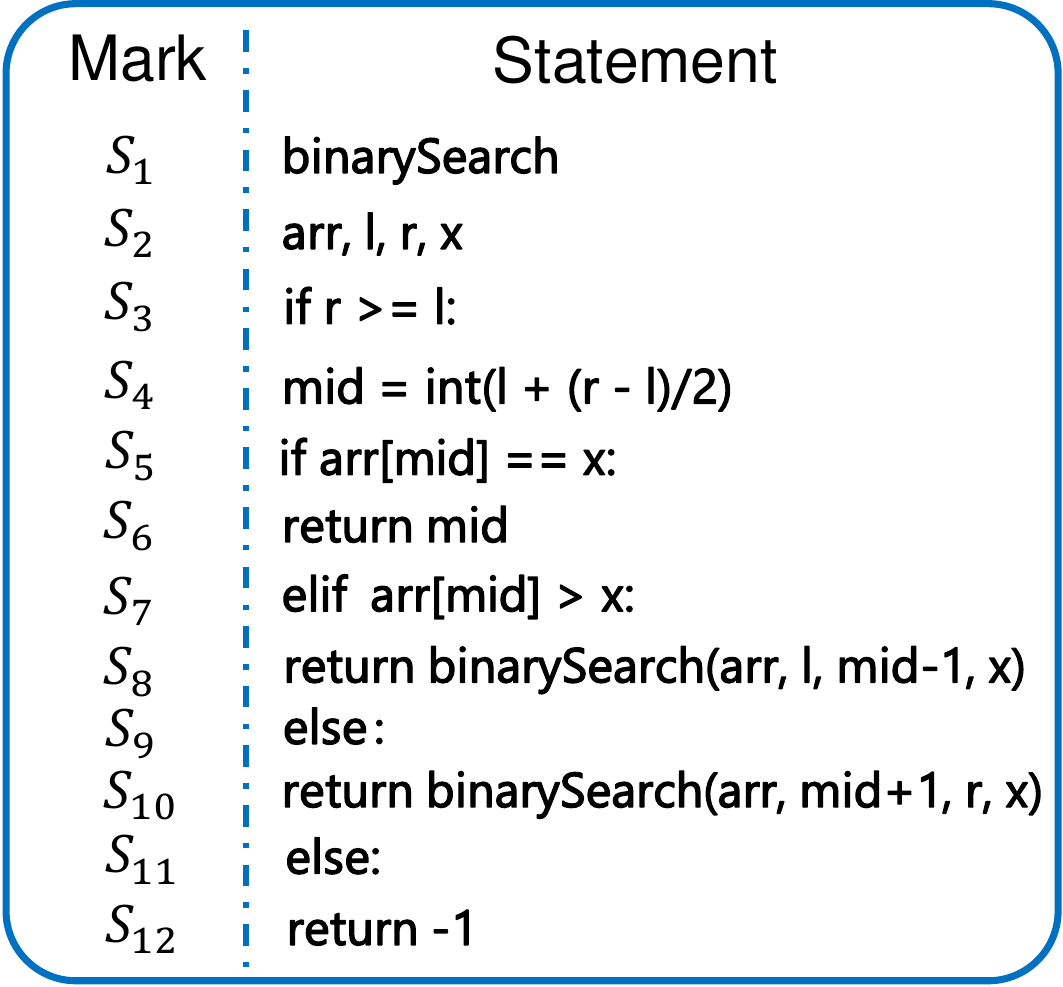} & \includegraphics[width=0.25 \textwidth]{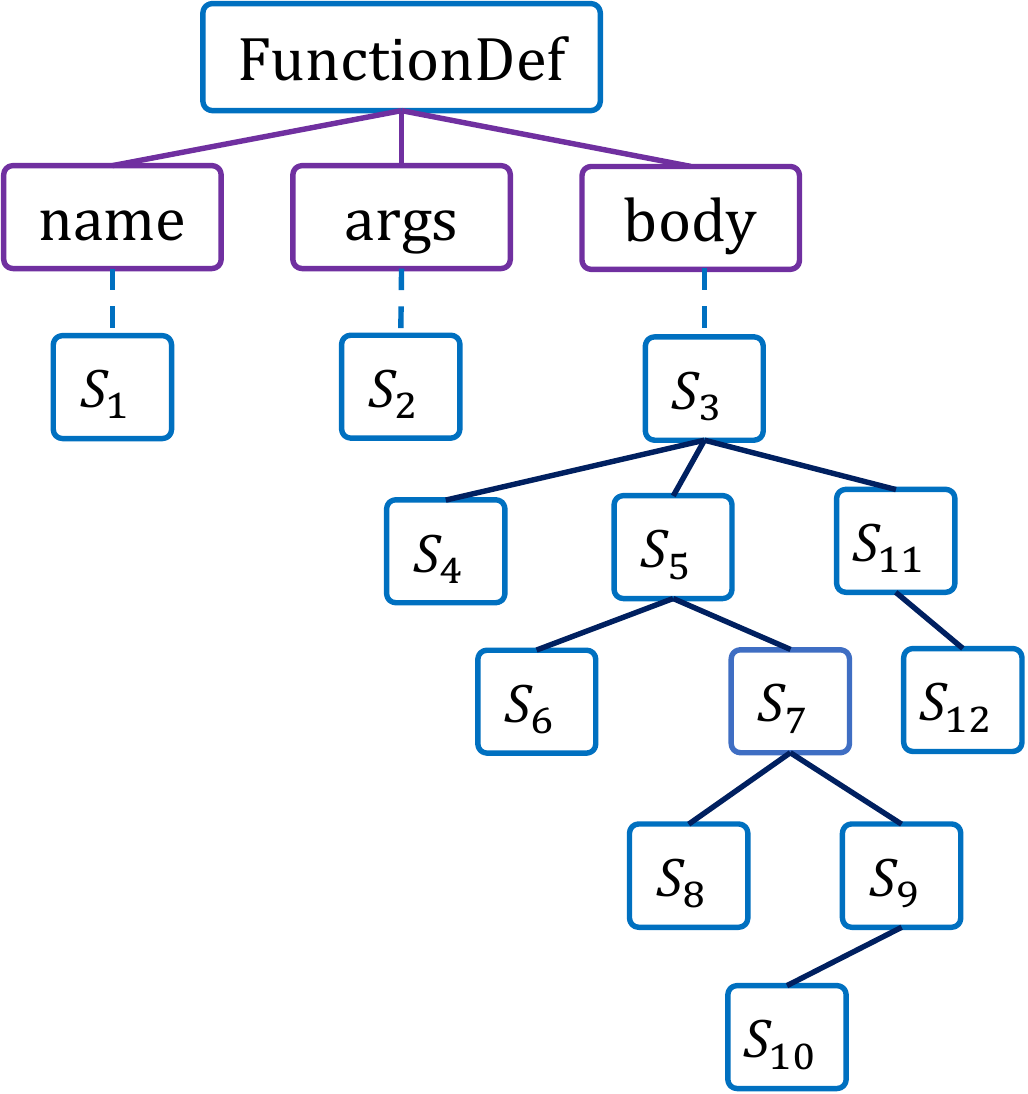} & \includegraphics[width=0.38\textwidth]{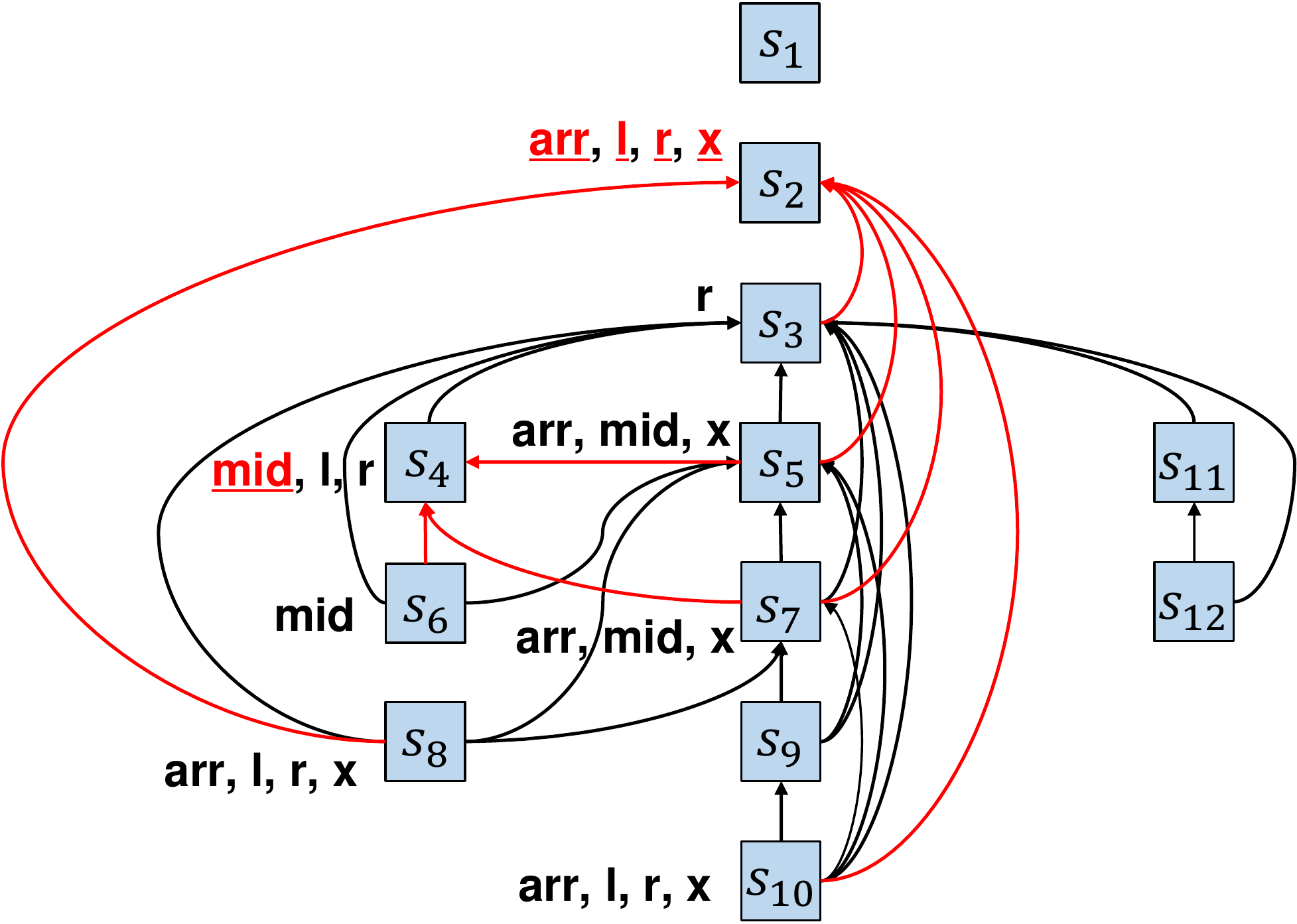} \\
	\small{(a) Marked statements.} & \small{(b) The simplified AST.} & \small{(c) The extracted PDG.}\\
	\end{tabular}
\caption{Workflow for extracting PDG of the code snippet in Listing~\ref{lst:example}. For the extracted PDG in (c), red and black arrowed lines indicate data dependency and control dependency respectively. The tokens beside each statement block denote the variables (re)defined (highlighted in red underlined font) or used in the corresponding statement.}
	\label{fig:dependency_graph}
\end{figure*}

For clarifying the PDG establishment process, we use the code example illustrated in Listing~\ref{lst:example}. Figure~\ref{fig:dependency_graph} (a) depicts the mark for each statement in the code example, in which we regard the function name and required parameters as two separate statements. Function name can be treated as a short summary of the code functionality; while the definitions of the required parameters generally reflect the semantics of the input data. Treating function names and parameters separately could be helpful for capturing their respective semantics.

Figure~\ref{fig:dependency_graph} (b) demonstrates the simplified AST of the code example where we construct the AST in statement level and hide the details of each statement. The \textit{data dependency} of one statement with the other statement can be identified if the variable used in one statement is (re)defined in the other statement and the value of the variable is unchanged on the execution path between these two statements. The \textit{control dependency} of one statement with the other is determined if the execution of the statement relies on the execution results of the other one. The control dependency can be directly captured by the tree structure in the AST, i.e., statements in child leaf nodes are considered possessing dependent relations with the statements in the parent nodes. The extracted PDG is depicted in Figure~\ref{fig:dependency_graph} (c), with red arrowed lines and black arrowed lines indicate data dependency and control dependency between the two statements, respectively. Tokens beside each statement block denote the related variables, in which we use black or red underlined variables to distinguish whether the variables are used or (re)defined in the corresponding statement. For example, the parent nodes of $S_{10}$ in the AST include $S_3$, $S_5$, $S_7$, and $S_9$ (as shown in Figure~\ref{fig:dependency_graph} (b)), so the control dependency between $S_{10}$ and $S_{3,5,7,9}$ is marked in the obtained PDG. Also, the variable \codeword{mid} in $S_6$, corresponding to line 5 in the Listing~\ref{lst:example}, is from $S_4$, i.e., line 3 in the code example; so $S_6$ shows a data dependency relation to $S_4$ in the PDG.

\subsubsection{Statement-Level Dependency Embedding}

The dependency embedding network is designed to encode the data dependency and control dependency involved in the PDG of a code snippet into a vector representation. According to the extracted PDG (as shown in Figure~\ref{fig:dependency_graph} (c)), we can build a dependency matrix $\Upsilon\in \{0,1\}^{(l)\times(l)}$, where $l$ indicates the number of statements in the code. The element $\upsilon_{ij}=1$ if the $i$-th statement has a data/control dependency on the $j$-th statement; otherwise $\upsilon_{ij}=0$. Note that $\upsilon_{ij}\neq\upsilon_{ji}$. For example, $S_4$ and $S_6$ exhibit a data dependency relation, so $\upsilon_{64}=1$. To embed the obtained dependency matrix $\Upsilon$, we employ one layer of multi-layer perceptron (MLP):

\begin{equation}
\begin{split}
    \mathbf{p}_i & = \text{tanh}(\mathbf{W}^{\Gamma}\mathbf{\upsilon}_i), \forall i=1,2,...,l,\\
    \mathbf{P} & = [\mathbf{p}_1, ..., \mathbf{p}_{(l)}],
    \label{con:dependencyembedding}
\end{split}
\end{equation}

\noindent where $\mathbf{W}^{\Gamma}$ is the matrix of trainable parameters in MLP and $\mathbf{p}_i$ is the embedding of the dependency information for each statement.

\subsubsection{Statement-Level Token Embedding}

The token embedding network is designed for capturing the semantics of each statement based on the constituted tokens. We first tokenize the statements into sequences of tokens following Gu et al.'s work~\citep{DBLP:conf/icse/GuZ018}, during which process duplicate tokens and the keywords in the programming language such as \codeword{while} and \codeword{break} are removed. Then tokens in each sequence are embedded into vectors individually through an embedding layer. An attention layer is utilized to compute a weighted average. Given a sequence of token embedding vectors $\{\mathbf{e}_{i,1},...,\mathbf{e}_{i,j},...\}$ for the $i$-th statement, the attention weight $\alpha_{i,j}$ for each $\mathbf{e}_{i,j}$ is calculated as follows:

\begin{equation}
    \alpha_{i,j}=\frac{\exp(\mathbf{e}^\intercal_{i,j})}{\sum_{j} \exp(\mathbf{e}^\intercal_{i,j})}.
    \label{con:attn1}
\end{equation}

\noindent Each statement is embedded based on the attention weights $\alpha_{i,j}$.

\begin{equation}
    \mathbf{t}_i = \sum_{j}\alpha_{i,j}e^\intercal_{i,j},
    \label{con:attn2}
\end{equation}

\noindent where $i$ indicates the $i$-th statement.

\subsubsection{Semantic Dependency Embedding}
We consider both statement-level dependency and semantic information for learning the vector representation of a code snippet. Specifically, for each statement $s_i$, we concatenate its dependency embedding $\mathbf{p}_i$ and token embedding $\mathbf{t}_i$ as the representation of the statement, i.e., $\mathbf{s}_i=[\mathbf{t}_i;\mathbf{p}_i]$. We finally adopt bi-LSTM to encode the sequence of the statement embeddings and use the last hidden state as the vector representation of the code.

\begin{equation}
   \mathbf{c} = \text{BiLSTM}(\mathbf{h}_{l},\mathbf{s}_{l}),
   \label{con:lstm}
\end{equation}

\noindent where $l$ indicates the number of statements.

\begin{mintedbox}[fontsize=\scriptsize]{python}
def binarySearch (arr, l, r, x): 
    if r >= l: 
        mid = int(l + (r - l)/2)
        if arr[mid] == x: 
            return mid 
        elif arr[mid] > x: 
            return binarySearch(arr, l, mid-1, x) 
        else: 
            return binarySearch(arr, mid+1, r, x) 
    else: 
        return -1
\end{mintedbox}
\begin{lstlisting}[frame=none,caption={An example of Python code snippet for illustrating the semantic dependency learning process.},captionpos=b,label=lst:example]
\end{lstlisting}

\subsection{Description Encoder}
The description encoder aims at embedding natural language descriptions into vectors. Given a description $D = \{w_1,...,w_{k},...,w_{N_d}\}$ comprising a sequence of $N_d$ words, the description encoder embeds it into a vector $\mathbf{d}$ using a bi-LSTM model with maxpooling:

\begin{equation}
\begin{split}
   \mathbf{h}_k &= \text{BiLSTM}(\mathbf{h}_{k-1},\mathbf{w}_k), \forall k=1,2,...,N_d,\\
   \mathbf{d} &= \text{maxpooling}([\mathbf{h}_1,...,\mathbf{h}_{N_d}]).
\end{split}
\end{equation}

The maxpooling layer is used to mitigate the effect of long-term information loss caused by the LSTM mechanism and catch the global feature of the whole sentence.

\subsection{Similarity Measurement}
The semantic similarity between the code vector $\mathbf{c}$ and description vector $\mathbf{d}$ is calculated based on its cosine distance in the embedding space:

\begin{equation}
   \text{cos}(\mathbf{c},\mathbf{d}) = \frac{\mathbf{c}^\intercal\mathbf{d}}{||\mathbf{c}||||\mathbf{d}||}.
   \label{con:lossfunction}
\end{equation}

The vector features of the two different embedding models are trained using the loss function, i.e., Equ.~\ref{con:lossfunction}, to maximize the cosine similarities in the projected space, so aligned code and descriptions would be close to each other in the space. Such design is widely adopted in prior code search studies~\cite{DBLP:conf/icse/GuZ018, DBLP:conf/sigsoft/CambroneroLKS019, DBLP:conf/pldi/SachdevLLKS018}. The target of the design is to get unified representations for both code and description, so as to mitigate the problem of semantic gap between them. The higher the similarity, the more relevant the code is to the description.

\subsection{Model Training}
We obtain the representation vectors for code snippets and descriptions based on the proposed code encoder and description encoder, respectively. Following previous studies~\cite{DBLP:conf/icse/GuZ018, DBLP:conf/sigsoft/CambroneroLKS019, DBLP:conf/pldi/SachdevLLKS018}, we project the code vectors and description vectors to the same space, and train the vectors for aligned code snippets and descriptions to be close in the space.

Specifically, every single code snippet in the training data $T$ will be constructed as a triplet $<C,D+,D->$. C represents the code snippet from the training Corpora, $D+$ indicates the description which semantically matches the code snippet in the ground truth, and $D-$ denotes the negative description which is randomly chosen from the training corpora with the true description excluded. The loss function is as below:

\begin{equation}
   \mathcal{L}(\theta)=\sum_{<C,D+,D-> \in T}\max(0, \epsilon- \text{cos}(\mathbf{c},\mathbf{d}\text{+}) + \text{cos}(\mathbf{c},\mathbf{d}\text{-})),
\end{equation}

\noindent where $\theta$ denotes the parameters in the proposed model, $\mathbf{c}$ denotes the code vector of $C$, $\mathbf{d}\text{+}$ and $\mathbf{d}\text{-}$ denote the description vectors of $D+$ and $D-$, respectively. Based on the training loss function, we can get unified representations for both code and description, thus mitigating the semantic gap between them.
\section{Experimental Setup}\label{sec:eval}
In the section, we introduce the collected dataset for experimentation, the evaluation metrics, implementation details and baseline models.

\subsection{Dataset Collection}
Two datasets are adopted for our experimental evaluation. One dataset is obtained from CodeSearchNet~\cite{DBLP:journals/corr/abs-1909-09436}, a publicly-available GitHub repository. We focus on the Python program language since it is one of the most popular programming languages, accounting for more than 30\% of the total market share as PYPL reported~\citep{pypl}. Detailed statistics of the dataset can be found in Table~\ref{tab:codesearchnet}. All the code in the corpus is in Python and with English descriptions. We have 407,126, 22,302, and 21,902 \textless code, description\textgreater\, pairs for training, validating and testing, respectively. The median and average numbers of the statements in the code are around 10. We also observe that the statements contain around three tokens on average, with the minimum at zero which is because the input parameters beside the method name are treated as an individual statement and some code snippets may not require any input parameters. Another dataset is from Code2seq~\citep{DBLP:conf/iclr/AlonBLY19}, with the statistics illustrated in Table~\ref{tab:code2seq}. We only select the code written in Python 3 from both datasets since the PDG extraction tool (introduced in Section~\ref{sec:pdg}) is specifically designed for Python 3 and may fail to parse the code written in Python 2.

Table~\ref{tab:dataset1} and Table~\ref{tab:dataset2} illustrate the distribution of statements numbers of the codes in the two dataset, i.e., CodeSearchNet and Code2Seq, respectively. We can observe that the long tail phenomenon occurs in the two datasets. Besides, more than 50\% of the code has $\leq$10 statements and more than 80\% has $\leq$20 statements.
\begin{table}[h]
\centering
\caption{Statistics of the number of statements in CodeSearchNet dataset.}
\label{tab:dataset1}
\begin{tabular}{crrr}
\toprule
\#Statements &Training Set&Validation Set&Test Set\\
\midrule
\midrule
0 $\sim$ 10  &230,183&12,413&12,326\\
11 $\sim$ 20 &117,060&6,364&6,361\\
21 $\sim$ 30 &32,904&1,875&1,843\\
31 $\sim$ 40 &12,834&755&633\\
41 $\sim$ 50 &5,723&386&326\\
51 $\sim$  &8,422&509&413\\
\midrule
\end{tabular}
\end{table}

\begin{table}[h]
\centering
\caption{Statistics of the number of statements in Code2Seq dataset.}
\label{tab:dataset2}
\begin{tabular}{crrr}
\toprule
\#Statements&Training Set&Validation Set &Test Set\\
\midrule
\midrule
0 $\sim$ 10  &218,679&32,429&33,210\\
11 $\sim$ 20 &73,870&11,301&12,251\\
21 $\sim$ 30 &2,0956&3,215&3,478\\
31 $\sim$ 40 &7,957&1,251&1,370\\
41 $\sim$ 50 &3,540&573&632\\
51 $\sim$  &4,326&650&786\\
\midrule
\end{tabular}
\end{table}

\begin{table}[h]
\centering
\caption{Statistics of the CodeSearchNet dataset.}
\label{tab:codesearchnet}
\begin{tabular}{lrrr}
\toprule
&Training&Validating&Testing\\
\midrule
\midrule
\# \textless code, description\textgreater &407,126&22,302&21,902\\
\midrule
\multicolumn{4}{c}{Statistics of \# statements in code}\\
Min.  &1&1&1\\
Med. &7&8&7\\
Max. &1,385&909&363\\
Ave.&11.45&11.87&11.24\\
\midrule
\multicolumn{4}{c}{Statistics of \# tokens in the statements}\\
Min. &0&0&0\\
Med. &3&3&3\\
Max. &514&155&83\\
Ave. &3.92&3.87&3.91\\
\bottomrule
\end{tabular}
\end{table}

\begin{table}[h]
\centering
\caption{Statistics of the Code2seq dataset.}
\label{tab:code2seq}
\begin{tabular}{lrrr}
\toprule
&Training&Validating&Testing\\
\midrule
\midrule
\# \textless code, description\textgreater &329,328&49,419&51,727\\
\midrule
\multicolumn{4}{c}{Statistics of \# statements in code}\\
Min.  &2&2&2\\
Med. &7&7&7\\
Max. &1,463&416&1,463\\
Ave.&10.17&10.33&10.68\\
\midrule
\multicolumn{4}{c}{Statistics of \# tokens in the statements}\\
Min. &0&0&0\\
Med. &3&3&3\\
Max. &682&199&1864\\
Ave. &3.75&3.73&3.73\\
\bottomrule
\end{tabular}
\end{table}

\subsection{Performance Measurement}
Following the evaluation settings in \citep{DBLP:conf/kbse/WanSSXZ0Y19}, we fix a set of 999 distractor snippets $\mathbf{c_j}$ for each test pair $(\mathbf{c_i},\mathbf{d_i})$ and calculate the average ranking score for all the testing pairs as the evaluation result. We involve two metrics: $R@k$ and MRR, for validating the ranking performance.  
\subsubsection{R@k} 
$R@k$ is a common metric to evaluate whether an approach can retrieve the correct answer in the top $k$ returning results. It is widely used by many studies on the code retrieval task. The metric is calculated as follows:
\begin{equation}
    R@k = \frac{1}{|Q|}\sum^{|Q|}_{q=1}\delta(FRank_q<k),
\end{equation}

\noindent where $Q$ denotes the query set and $FRank_q$ denotes the rank of the correct answer for query $q$. The function $\delta(Frank_q<k)$ returns 1 if the rank of the correct answer within the top $k$ returning results otherwise the function returns 0. A higher $R@k$ indicates a better code retrieval performance.

\subsubsection{MRR}
Mean Reciprocal Rank (MRR) is the average of the reciprocal ranks of the correct answers of query set $Q$, which is another popular evaluation metric for the code retrieval task. The metric MRR is calculated as follows:

\begin{equation}
    MRR = \frac{1}{|Q|}\sum^{|Q|}_{q=1}\frac{1}{FRank_q}.
\end{equation}

The higher the MRR value is, the better performance the model has.

\subsection{Implementation Details}
In our experiment, we select the top 10,000 words according to the word frequencies as the vocabularies of code snippets and descriptions, respectively. All the word embeddings are randomly initialized and adjusted during training. The dimension of word embedding is set as 256. All LSTMs have 1024 hidden units in each direction. The maximum number of considered statements in the code and the maximum number of tokens in each statement are set as 20 and 5, respectively. The sequence lengths of descriptions are limited as 30 following the work~\citep{DBLP:conf/icse/GuZ018}. The \tool model is trained via the AdamW algorithm\citep{DBLP:journals/corr/KingmaB14} and the learning rate is 2.08$e$-4. To mitigate the over-fitting issue, we add a dropout layer with dropout rate at 0.25. We train our models on a server with one Nvidia GeForce RTX 2080 Ti and 11 GB memory. The training lasts $\sim$20 hours with 200 epochs and the early stopping strategy~\citep{goodfellow2016deep} is adopted to avoid overfitting.

\subsection{Baseline Models}
We compare our proposed model with several state-of-the-art baseline models. \textbf{CODEnn} is one of the state-of-the-art models proposed in~\citep{DBLP:conf/icse/GuZ018}. This model extracts the method name, API sequence and tokens from the code and utilizes neural network to learn the unified vector representation of query and these code features. \textbf{UNIF}~\cite{DBLP:conf/sigsoft/CambroneroLKS019} focuses on the semantic information from the tokens in the code and utilizes embedding techniques and attention mechanism to embed the tokens in the query and code into a single vector respectively. The projection of the query and code vector in the same space is learned by this model. \textbf{NeuralBoW}~\citep{DBLP:conf/acl/WangM12} embeds each token in the two input sequences to a learnable embedding. The token embeddings are then combined into a sequence embedding using max-pooling and an attention-like weighted sum mechanism. The \textbf{RNN} baseline adopts two-layer bi-directional LSTM model~\citep{DBLP:conf/ssst/ChoMBB14} to encode the input sequences.
% , with the number of hidden units and dropout rate defined at 64 and 0.8, respectively.
\textbf{CONV}~\citep{DBLP:conf/emnlp/Kim14} uses 1D convolutional neural network over both the input sequences of tokens. \textbf{CONVSelf}~\citep{DBLP:conf/iclr/LinFSYXZB17} combines 1D convolutional neural network and self-attention layer to embed both input sequences. \textbf{SelfAttn}~\citep{DBLP:journals/corr/abs-1909-09436} utilizes multi-head attention~\cite{DBLP:conf/nips/VaswaniSPUJGKP17} to encode both input sequences of tokens, and has proven effective on multiple types of programming languages such as Python and JavaScript. The hyper-parameters of the baselines are defined according to the original papers~\citep{DBLP:conf/icse/GuZ018,DBLP:conf/sigsoft/CambroneroLKS019,DBLP:journals/corr/abs-1909-09436}. During implementing CODEnn, NeuralBoW, RNN, CONV, CONVSelf and SelfAttn, we directly utilized the released code; while for UNIF, we tried our best to replicate the code according to the paper and will make the replication publicly available.

\section{Experimental Results}\label{sec:result}
In this section, we present the evaluation results, including the main results, parameter analysis, case studies and error analysis.

\subsection{Main Results}

\textbf{Involving semantic dependency embeddings increases the code search performance.} Table~\ref{tab:results for codesearchnet} and \ref{tab:results for code2seq} illustrate the evaluation results comparing with the baseline models. As can be seen, \tool presents the best performance comparing with all the baseline models, increasing the performance of 36.38\% in terms of $R@1$, 17.13\% in term of $R@5$, 12.54\% in term of $R@10$ and 
25.26\% in term of $MRR$ at least on the dataset of CodeSearchNet. \tool can achieve the improvement of the performance at least 22.34\%, 22.51\%, 21.54\% and 21.79\% in $R@1$, $R@5$, $R@10$ and $MRR$ on the dataset of Code2Seq, respectively.
This indicates that \tool can rank the correct answer the top more accurately when given a natural language query. The improvement on $R@1$ is most significant among all the metrics in our proposed model, which is over 20\% in both datasets. $R@1$ is the metric concerned most by programmers since they prefer to use the code search system which can return the best results in first. The higher $MRR$ score further verifies the effectiveness of \tool.
The difference between \tool and the baseline models is the code representation strategy, which shows the effectiveness of the semantic dependency embeddings for code search.

\textbf{Attention mechanism can be helpful for effective code search.} By comparing CONV with CONVSelf, we can observe that with the attention mechanism integrated, CONV presents a better performance than the pure CONV model on both datasets. For example, CONVSelf increases the accuracy of CONV by 20.21\% and 15.37\% in terms of $R@1$ and $MRR$ on the CodeSearchNet dataset, respectively. Similar result also appears on the Code2Seq dataset. The results imply the effectiveness of the attention mechanism on the code search task. We also compared with the performance of the \tool\textsubscript{maxpooling} where the attention mechanism is replaced with the max pooling strategy~\citep{lee2016generalizing}. As can be seen in Table~\ref{tab:results for codesearchnet} and Table~\ref{tab:results for code2seq}, \tool with the attention mechanism involved outperforms the \tool with max pooling strategy integrated on both datasets, which further demonstrates the effectiveness of the attention mechanism on the task.

\textbf{\tool shows better generalizability than baseline models.} As can be observed from Table~\ref{tab:results for codesearchnet} and Table~\ref{tab:results for code2seq}, one baseline model's extraordinary performance on a specific dataset can not transfer to other datasets. For example, SelfAttn achieves the best performance among all the baselines on the CodeSearchNet dataset with respect to $R@1$, but perform worse than NeuralBoW on the Code2seq dataset. Comparing with the baselines, \tool presents the best performance on both datasets, which can explicate the good generalizability of \tool.

\begin{table}[t]
\centering
\caption{Comparison results with baseline models on the CodeSearchNet dataset. The best results are highlighted in \textbf{bold} fonts.}
\label{tab:results for codesearchnet}
\begin{tabular}{lcccc}
\toprule
Approach&R@1&R@5&R@10&MRR\\
\midrule
\midrule
CODEnn&0.367&0.573&0.652&0.465\\
UNIF&0.379&0.615&0.706&0.490\\
NeuralBoW&0.521&0.747&0.807&0.622\\
RNN&0.556&0.772&0.832&0.654\\
CONV&0.475&0.703&0.776&0.579\\
CONVSelf&0.571&0.788&0.845&0.668\\
SelfAttn&0.580&0.786&0.840&0.673\\
\midrule
\midrule
\tool\textsubscript{maxpooling} & 0.777 & 0.914 & 0.946 & 0.838 \\
\tool&\textbf{0.791}&\textbf{0.923}&\textbf{0.951}&\textbf{0.843}\\
\bottomrule
\end{tabular}
\end{table}

\begin{table}[t]
\centering
\caption{Comparison results with baseline models on the Code2seq dataset. The best results are highlighted in \textbf{bold} fonts.}
\label{tab:results for code2seq}
\begin{tabular}{lcccc}
\toprule
Approach&R@1&R@5&R@10&MRR\\
\midrule
\midrule
CODEnn&0.330&0.532&0.617&0.427\\
UNIF&0.380&0.588&0.668&0.478\\
NeuralBoW&0.546&0.693&0.738&0.615\\
RNN&0.438&0.623&0.688&0.526\\
CONV&0.425&0.584&0.645&0.502\\
CONVSelf&0.470&0.642&0.700&0.552\\
SelfAttn&0.525&0.683&0.731&0.599\\
\midrule
\midrule
\tool\textsubscript{maxpooling} &   0.664 &  0.843 & 0.892 & 0.745  \\
\tool&\textbf{0.668}&\textbf{0.849}&\textbf{0.897}&\textbf{0.749}\\
\bottomrule
\end{tabular}
\end{table}

\subsection{Parameter Analysis}

\begin{figure*}[ht]
\centering
\subfigure[\# hidden units in LSTMs]{
\begin{minipage}[t]{0.33\linewidth}
\centering
\includegraphics[width=5.8cm]{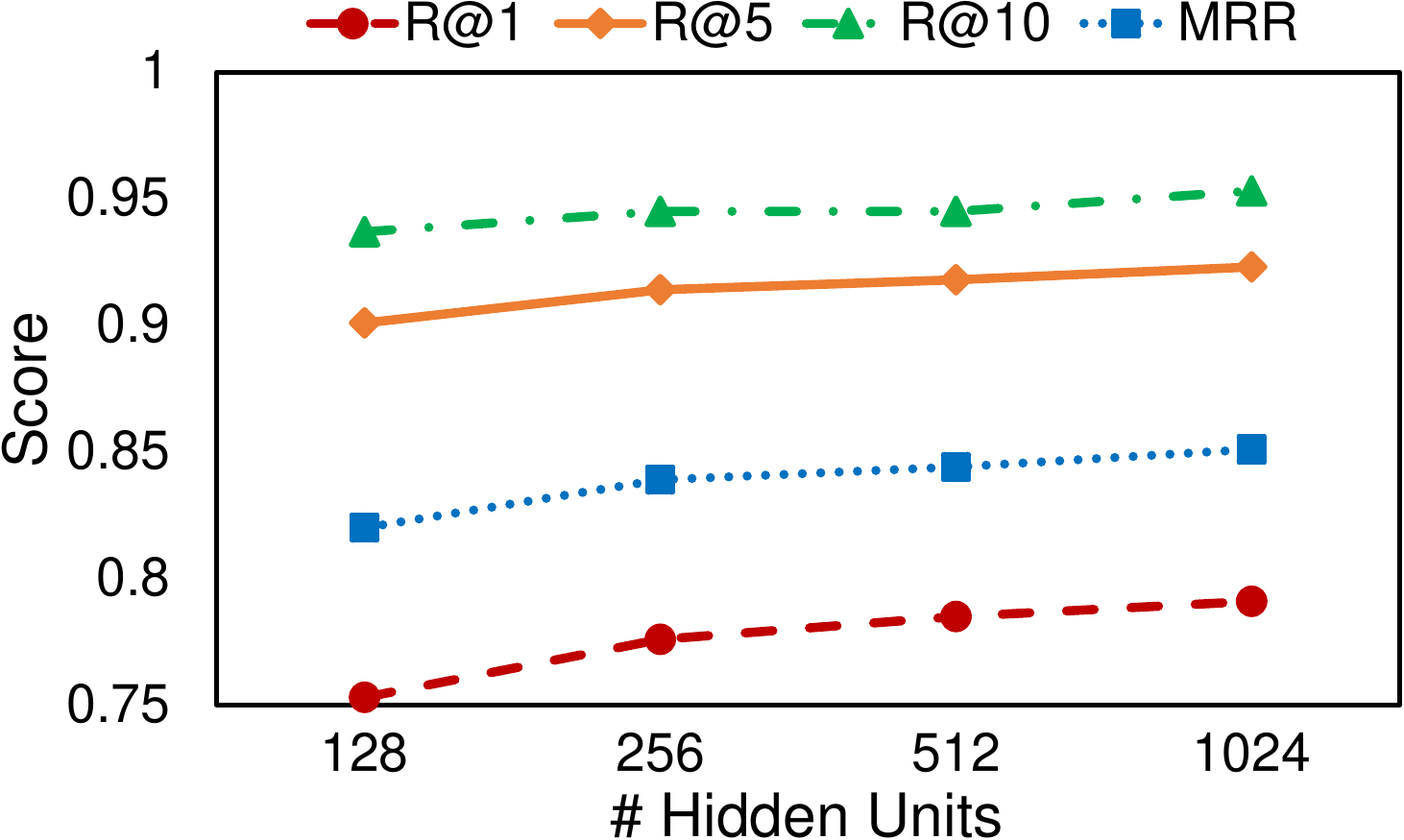}
\label{fig:hidden_size_cn}
\end{minipage}%
}%
\subfigure[\# maximum statements in code]{
\begin{minipage}[t]{0.33\linewidth}
\centering
\includegraphics[width=5.8cm]{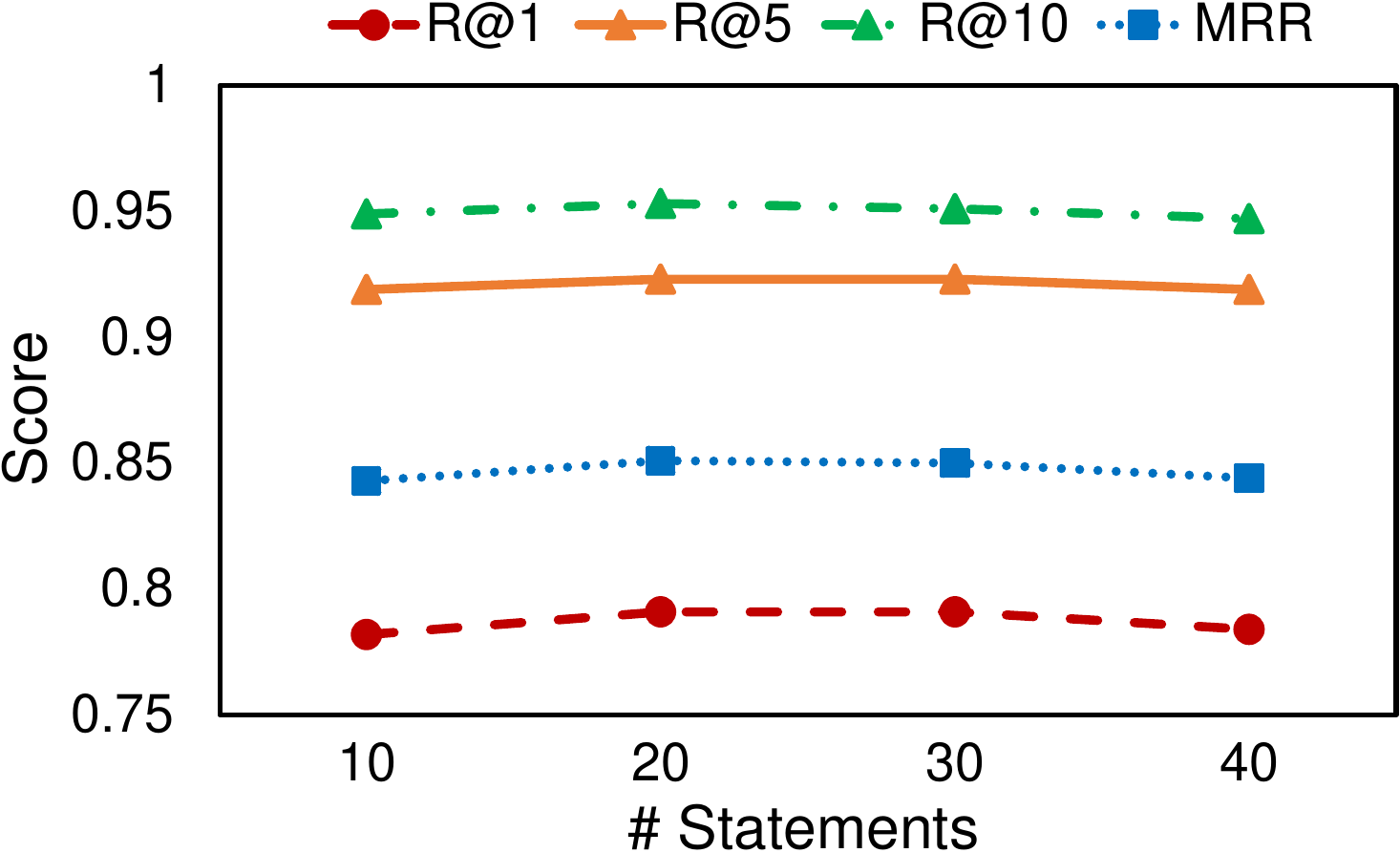}
\label{fig:statement_len_cn}
\end{minipage}%
}%
\subfigure[\# maximum tokens in statement]{
\begin{minipage}[t]{0.33\linewidth}
\centering
\includegraphics[width=5.8cm]{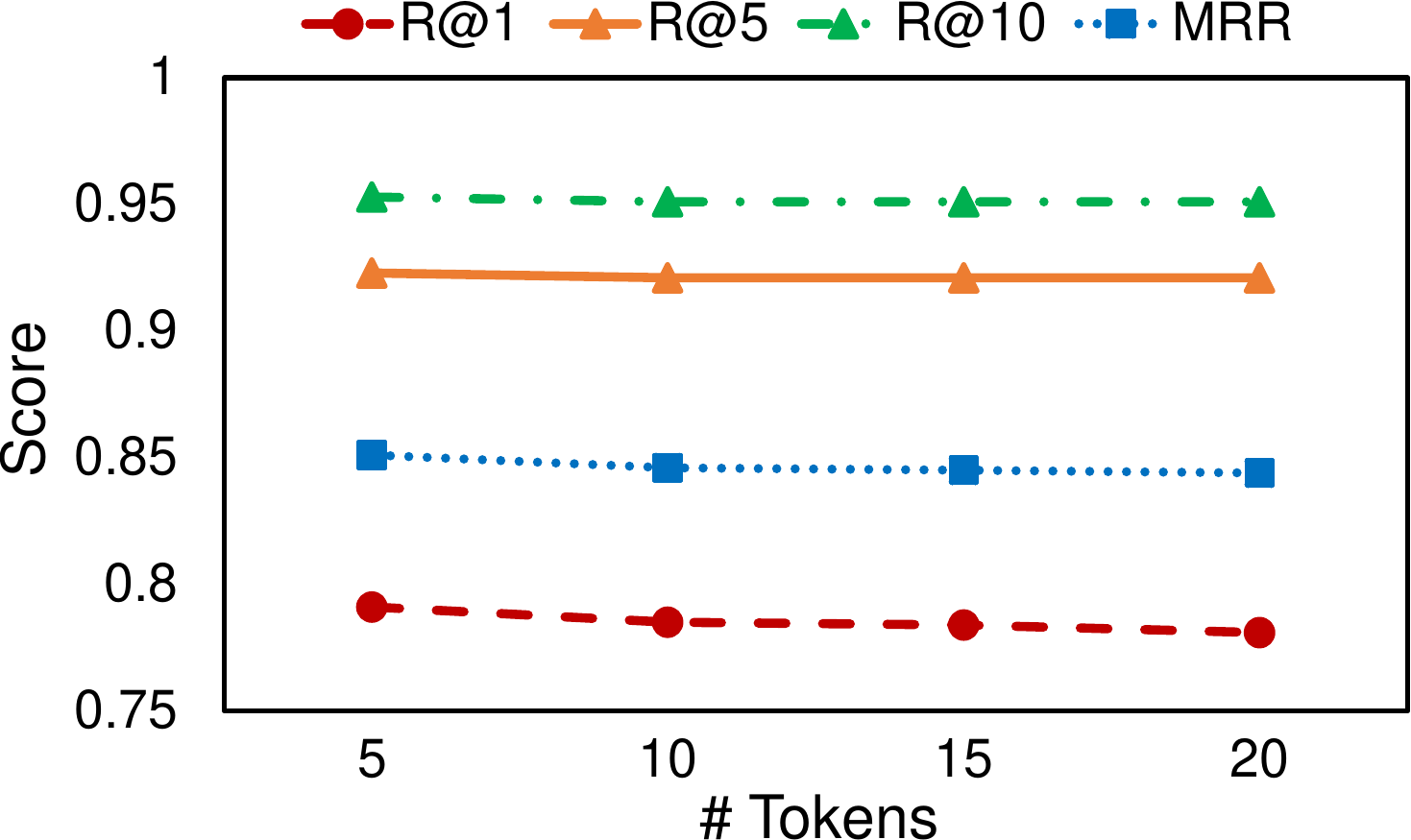}
\label{fig:token_len_cn}
\end{minipage}
}%
\centering
\caption{Parameter sensitivity study for CodeSearchNet.}
\label{fig:parameter_cn}
\end{figure*}

\begin{figure*}[ht]
\centering
\subfigure[\# hidden units in LSTMs]{
\begin{minipage}[t]{0.33\linewidth}
\centering
\includegraphics[width=5.8cm]{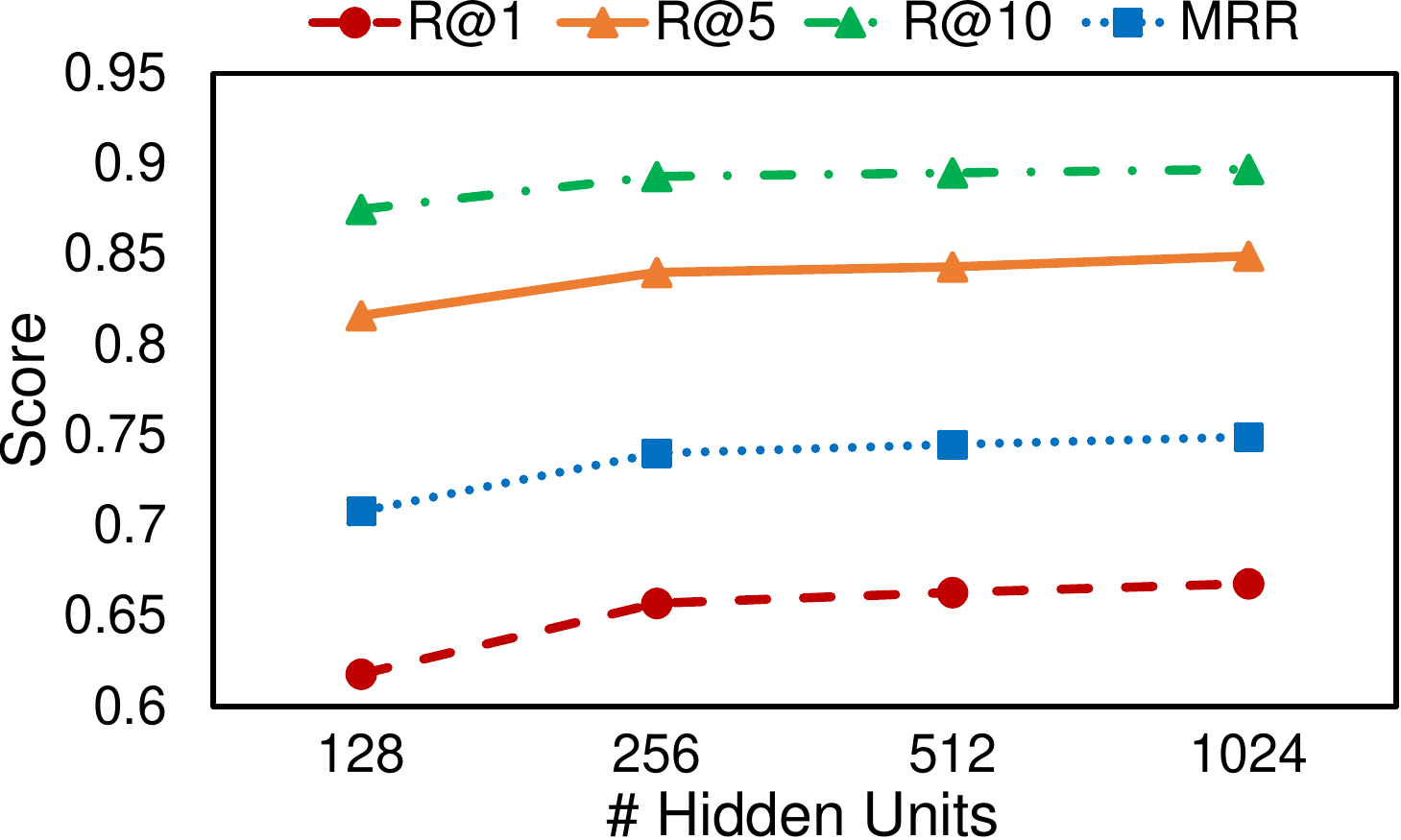}
\label{fig:hidden_size_cs}
\end{minipage}%
}%
\subfigure[\# maximum statements in code]{
\begin{minipage}[t]{0.33\linewidth}
\centering
\includegraphics[width=5.8cm]{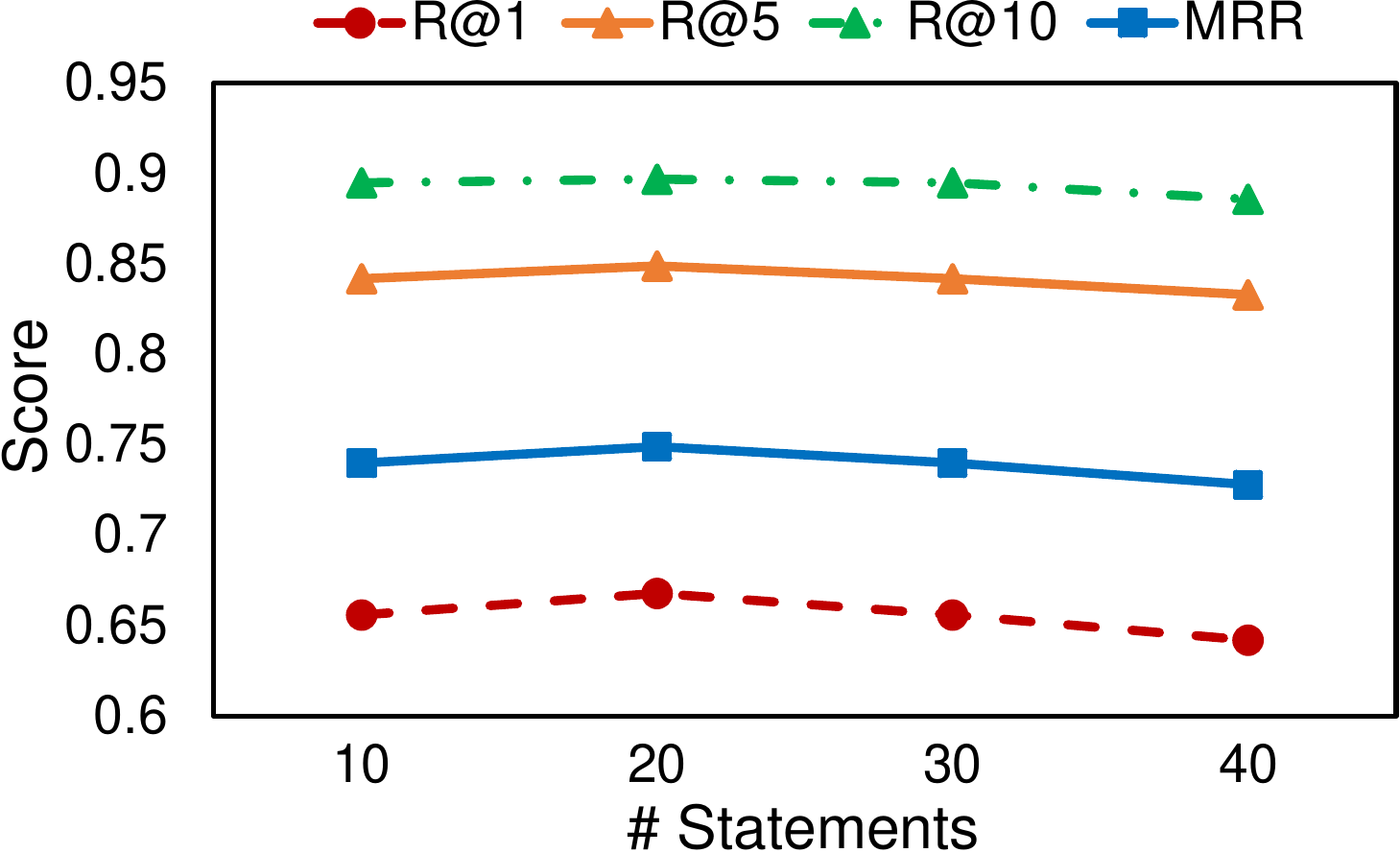}
\label{fig:statement_len_cs}
\end{minipage}%
}%
\subfigure[\# maximum tokens in statement]{
\begin{minipage}[t]{0.33\linewidth}
\centering
\includegraphics[width=5.8cm]{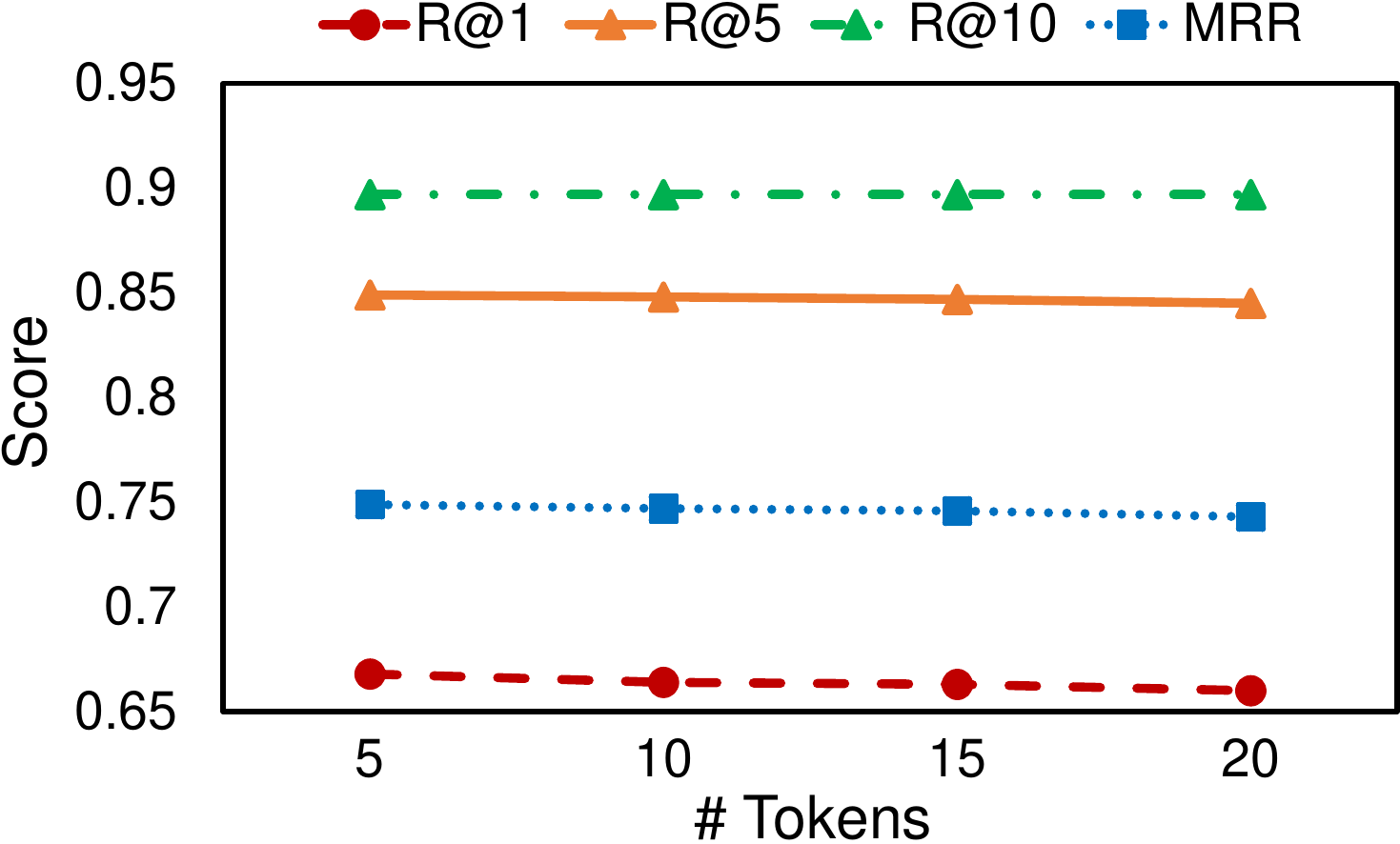}
\label{fig:token_len_cs}
\end{minipage}
}%
\centering
\caption{Parameter sensitivity study for Code2Seq.}
\label{fig:parameter_cs}
\end{figure*}

In this section, we will discuss how the hyperparameters affect the performance of \tool. Three hyperparameters are analyzed, including the number of hidden units in LSTMs, the maximum number of considered statements in the code, and the maximum number of considered tokens in each statement. Figure~\ref{fig:parameter_cn} and Figure~\ref{fig:parameter_cs} depict the results of the parameter analysis.

\subsubsection{\# Hidden units in LSTMs} As shown in Figure~\ref{fig:hidden_size_cn} and Figure~\ref{fig:hidden_size_cs}, all the metric values present an increasing trend as the number of hidden units grows. The phenomenon is understandable since more hidden units imply that the model has more parameters to learn and can extract more knowledge from the same input. We can also observe that for each doubling of the number of hidden units, the growth rates of the $R@1$ scores are 1.9\%, 0.54\%, 0.41\% respectively on the CodeSearchNet dataset. The trend is identical for the Code2Seq dataset. So we can summarize that with an increasing number of the hidden units, the model performance would increase but the increasing rates show a declining tendency. Due to the limitation of the computing source and the marginal enhancement when the number of hidden units is larger than 1,024, we choose 1,024 as the number of hidden units for our experiment.

\subsubsection{\# Maximum statements in code} Figure~\ref{fig:statement_len_cn} and Figure~\ref{fig:statement_len_cs} illustrate the variations of the model performance as the maximum number of considered statements increases. We can observe that the metrics achieve the highest values when the number equals 20 and manifests a declining trend as the statement number further increases. As can be found in Table~\ref{tab:codesearchnet} and Table~\ref{tab:code2seq}, the median numbers of the statements in both CodeSearchNet and Code2seq datasets are 7, with the average at around 10. Thus, more statements considered would not be beneficial for capturing the code semantics for most code snippets. In the experiment, we set the maximum number of considered statements in code as 20.

\subsubsection{\# Maximum tokens in statement} The impact of different maximum numbers of involved tokens in a statement is shown in Figure~\ref{fig:token_len_cn} and Figure~\ref{fig:token_len_cs}. We can find that when the involved token number increases, the performance presents a downward trend. According to Table~\ref{tab:codesearchnet} and Table~\ref{tab:code2seq}, the average number of tokens in the statements is $\sim$3. So with more tokens recognized, the model could not learn more knowledge of the code snippets. In the experiment, to balance the model performance with the number of tokens considered, we define the maximum number of the tokens in a statement as 5.

\begin{mintedbox}[fontsize=\scriptsize]{python}
def logs(self, prefix='worker'):
    logs = []
    logs += [('success_rate', np.mean(self.success_history))]
    if self.compute_Q:
        logs += [('mean_Q', np.mean(self.Q_history))]
    logs += [('episode', self.n_episodes)]

    if prefix != '' and not prefix.endswith('/'):
        return [(prefix + '/' + key, val) for key, val in logs]
    else:
        return logs
\end{mintedbox}
\begin{lstlisting}[frame=none,caption={Successful case 1.},captionpos=b,label=lst:case1]
\end{lstlisting}

\subsection{Ablation Study}
In the ablation study, we validate the contribution of data dependency or control dependency to \tool and the effectiveness of combing both dependency types.
Table~\ref{tab:ablation study for codesearchnet} and Table~\ref{tab:ablation study for code2seq} shows the results of the ablation study on the datasets of CodeSearchNet and Code2seq, respectively. $\text{CRaDLe}_\text{Full}$ represents the model utilizes both data dependency and control dependency, $\text{CRaDLe}_\text{DataDependency}$ represents the model only employs data dependency and $\text{CRaDLe}_\text{ControlDependency}$ represents the mode only utilizes control dependency.

From the results, we can find that the performance of the model that only utilizes data dependency is very close to the performance of the model with only control dependency, which shows that the importance of data dependency and control dependency is relatively equivalent under our implementation. However, we can find that the model that contains both data dependency and control dependency outperforms the model that only contains one dependency type, especially in terms of the R@1 metric. The results indicate that the combination of data dependency and control dependency is beneficial for effective code search.

\begin{table}[h]
\centering
\caption{Ablation study on the CodeSearchNet dataset.}
% \hy{it is surprising that plain RNN can do quite good}}
\label{tab:ablation study for codesearchnet}
\begin{tabular}{lcccc}
\toprule
Approach&R@1&R@5&R@10&MRR\\
\midrule
\midrule
$\text{CRaDLe}_\text{Full}$ & 0.791 & 0.923 & 0.951 & 0.843 \\
$\text{CRaDLe}_\text{DataDependency}$ & 0.779 & 0.910 & 0.946 & 0.840\\
$\text{CRaDLe}_\text{ControlDependency}$ & 0.785 & 0.918 &0.950 &0.845\\
\bottomrule
\end{tabular}
\end{table}

\begin{table}[h]
\centering
\caption{Ablation study on the Code2seq dataset.}
\label{tab:ablation study for code2seq}
\begin{tabular}{lcccc}
\toprule
Approach&R@1&R@5&R@10&MRR\\
\midrule
\midrule
$\text{CRaDLe}_\text{Full}$& 0.668 & 0.849 & 0.897 & 0.749\\
$\text{CRaDLe}_\text{DataDependency}$& 0.645 & 0.827 & 0.880 & 0.724\\
$\text{CRaDLe}_\text{ControlDependency}$& 0.645 & 0.828 & 0.882 & 0.730\\
\bottomrule
\end{tabular}
\end{table}

\subsection{Case Studies}

Listing~\ref{lst:case1} shows our predicted code snippet for the query ``\textit{Generates a dictionary that contains all collected statistics}''. We can find that our predicted result correctly matches the given query. Although no overlapping words exist between the code and query, \tool could capture that the code tokens such as \codeword{rate} and \codeword{compute} are semantically related to the query word ``\textit{statistics}''. Besides, since the semantically-related tokens mainly appear at the line 3, 4 and 5, and do not span the entire code, we guess that the involved dependency information helps to establish the relationships among the statements.

Listing~\ref{lst:case2} shows another predicted code snippet that accurately matches the given query ``\textit{Tile N images into one big PxQ image (P,Q)}''. Clearly, the function name contains the keywords in the query, e.g., ``\textit{tile}" and ``\textit{images}". Moreover, the core idea of this query is to tile N images into one image, essentially related to matrix operations. As shown in the Listing~\ref{lst:case2}, the code contains tokens associated with matrix transformation such as \codeword{reshape} and \codeword{transpose}. So with statement-level tokens explicitly incorporated, \tool could well catch the code functionality.

\begin{mintedbox}[fontsize=\scriptsize]{python}
def tile_images(img_nhwc):
    img_nhwc = np.asarray(img_nhwc)
    N, h, w, c = img_nhwc.shape
    H = int(np.ceil(np.sqrt(N)))
    W = int(np.ceil(float(N)/H))
    img_nhwc = np.array(list(img_nhwc) + [img_nhwc[0]*0 for _ in range(N, H*W)])
    img_HWhwc = img_nhwc.reshape(H, W, h, w, c)
    img_HhWwc = img_HWhwc.transpose(0, 2, 1, 3, 4)
    img_Hh_Ww_c = img_HhWwc.reshape(H*h, W*w, c)
    return img_Hh_Ww_c
\end{mintedbox}
\begin{lstlisting}[frame=none,caption={Successful case 2.},captionpos=b,label=lst:case2]
\end{lstlisting}

Overall, the above two examples indicate that \tool can accurately capture the code semantics with the statement-level dependency and semantic information integrated.

\subsection{Error Analysis}
Although most of the time, our model returns correct code snippets, we still notice that our model fails under the following two particular circumstances.  

\subsubsection{Code Containing Complex Mathematical Logic}
Listing~\ref{lst:error1} provides a failure case where the code contains complex mathematical logic. The description corresponding to the code is ``\textit{Convert directly the matrix from Cartesian coordinates (the origin in the middle of image) to Image coordinates (the origin on the top-left of image)}", which includes some mathematical concepts such as ``\textit{Cartesian coordinates}''. Nevertheless, no words related to the mathematical concepts appear in the code. Less knowledge learnt about the mathematical terminology renders the model harder to capture the semantic relevance between the code and natural language. Future work can incorporate external knowledge such as API documentation or Wikipedia for enhancing the understanding of the mathematical concepts.

\begin{mintedbox}[fontsize=\scriptsize]{python}
def transform_matrix_offset_center(matrix, y, x):
    o_x = (x - 1) / 2.0
    o_y = (y - 1) / 2.0
    offset_matrix = np.array([[1, 0, o_x], [0, 1, o_y], [0, 0, 1]])
    reset_matrix = np.array([[1, 0, -o_x], [0, 1, -o_y], [0, 0, 1]])
    transform_matrix = np.dot(np.dot(offset_matrix, matrix), reset_matrix)
    return transform_matrix

\end{mintedbox}
\begin{lstlisting}[frame=none,caption={Failure case 1.},captionpos=b,label=lst:error1]
\end{lstlisting}

\subsubsection{Code Containing Function Invocation} %ircuitous Execution Logic
We also find that the proposed model may fail to capture the code semantics when the code involves function invocation but the details of the invoked function are missing. Listing~\ref{lst:error2} illustrates such an example, and the corresponding description is ``\textit{Get successor to key, raises KeyError if a key is max key or key does not exist}''. As can be seen in the code example, the execution results strongly rely on the invoked function \codeword{succ\_item()}, however, the implementation of the invoked function is not detailed. For the case, the code semantics is difficult to be fully captured by the model, leading to failure.

\begin{mintedbox}[fontsize=\scriptsize]{python}
def succ_key(self, key, default=_sentinel):
        item = self.succ_item(key, default)
        return default if item is default else item[0]
\end{mintedbox}
\begin{lstlisting}[frame=none,caption={Failure case 2.},captionpos=b,label=lst:error2]
\end{lstlisting}

\section{Discussion}
\subsection{Dependency Embedding Approach}
In this section, we design another method for representing the dependency information. Specifically, we enrich the dependency matrix with the semantics of the tokens at statement level. The statement-level dependency embedding is calculated as below:
\begin{equation}
\begin{split}
    \mathbf{p}_i & =\frac{\sum_j{\mathbf{t}_j\mathbf{\upsilon}_{ij}}}{\max{(1, \sum_j{\mathbf{\upsilon}_{ij}}})}, \forall i=1,2,...,l,\\
    \mathbf{P} & = [\mathbf{p}_1, ..., \mathbf{p}_{(l)}],
\end{split}
\end{equation}
\noindent where $\mathbf{t}_j$ represents the statement-level token embedding for $j$-th statement, which is calculated via Equ.~\ref{con:dependencyembedding}. $\upsilon_{ij}$ indicates whether the $i$-th statement has a data/control dependency on the $j$-th statement and $\mathbf{p}_i$ is the new dependency embedding.

We evaluated the performance of new dependency embedding methods on the datasets of CodeSearchNet and Cod\-e2seq, as shown in Table~\ref{tab:results for new implementation}. 

From the table, we can find that the new strategy for encoding the dependency information outperforms our original approach in terms of the R@1 and MRR metrics for both datasets. The results indicate that the new approach for the dependency embedding may be more effective than the original approach for the task.

Graph neural networks (GNNs) is also a potential way to represent the dependency between different statements in one code snippet. However, using GNNs for representing the semantic dependency of code is beyond the scope of the work, since the assumption of GNNs that adjacent nodes share similar semantics no long holds for the control dependency information, and it would be more challenging to encode the semantic dependency information through GNNs. In the future, we will investigate various strategies to embed the semantic dependency with GNNs~\cite{DBLP:journals/nn/LiMWZ20,DBLP:conf/icml/WangLMMZF20}.

\begin{table}[t]
\centering
\caption{Comparison results with our original models. The best results are highlighted in \textbf{bold} fonts.}
\label{tab:results for new implementation}
\setlength{\tabcolsep}{1mm}{
\begin{tabular}{clcccc}
\toprule
Dataset&Approach&R@1&R@5&R@10&MRR\\
\midrule
\midrule
\multirow{2}{*}{CodeSearchNet} &CRaDLe\textsubscript{original} & 0.791 & \textbf{0.923} & \textbf{0.951} & 0.843 \\
\cmidrule{2-6}
& CRaDLe\textsubscript{new} &\textbf{0.794} & 0.920 & 0.949 & \textbf{0.851}\\
\midrule
\midrule
\multirow{2}{*}{Code2seq} &CRaDLe\textsubscript{original} & 0.668 & 0.849 & 0.897 & 0.749\\
\cmidrule{2-6}
& CRaDLe\textsubscript{new} & \textbf{0.676} &\textbf{0.852}&\textbf{0.899}&\textbf{0.756}\\
\bottomrule
\end{tabular}
}
\end{table}

\section{Related Work}\label{sec:literature}
The work is inspired by the studies related to both code search and code semantics representation learning.

\subsection{Code Search}
In software development, developers accomplish the goal of effective and high quality code by reusing the existing huge amount of available code resources. Prior work has explored a number of methods to find the implicit connections between human language queries and code databases. Early studies concentrate mainly on extracting useful features from both codes and queries. For example, the work \citep{DBLP:conf/aosd/ShepherdFHPV07} extracts scattered verbs from queries and applies action-oriented identiﬁer graph model to inspect the result graph, which helps to optimize the queries. In \citep{DBLP:conf/wcre/LuSWLD15}, Lu et al. reformulate and extract natural language phrases from source code identifiers since the synonyms in source codes and NL queries may affect the code search result significantly. The work \citep{DBLP:conf/icse/McMillanGPXF11} proposes Portfolio uses random surfer to model the navigation behavior of programmers. Then with association model based on Spreading Activation Network \citep{DBLP:journals/air/Crestani97}, functional relevant functions can be set in the same list. Ponzanelli et al.~\citep{DBLP:conf/msr/PonzanelliBPOL14} propose to retrieve pertinent discussions from Stack Overﬂow when given a context in the IDE, which saves developers' time spent on formulating more standardized queries.

With rapid development of deep learning, an increasing amount of work has focused on using neural networks for effective code search. In the work \citep{DBLP:conf/pldi/SachdevLLKS018}, Sachdev et al. first develop neural code search model called NCS to conduct NL search directly over large source code corpora. In Liu et al.' work \citep{DBLP:conf/pldi/LiuKMC019}, they present a neural model called NQE, which expands the queries and improves performance for shorter queries. Codenn embeds both code snippets and natural language descriptions into a unified vector space, in such way that code and its corresponding NL description have similar vectors \citep{DBLP:conf/icse/GuZ018}. Iyer et al. \citep{DBLP:conf/acl/IyerKCZ16} use attentional long short term memory (LSTM) networks to focus on more cardinal parts of the source code to produce search queries. Cambronero et al. propose UNIF, a bag-of-words-based network which includes API sequences, method tokens, method body tokens and docstring tokens for representing source code \citep{DBLP:conf/sigsoft/CambroneroLKS019}. Yao, Pedamail, and Sun regard code annotation and code search as dual task and consider the generated code annotations for better code search \citep{DBLP:conf/www/YaoPS19}. Husain et al. explore the semantic representations of different neural architectures and they find that self-attention-based architectures achieve the best performance~\citep{DBLP:journals/corr/abs-1909-09436}.

\subsection{Representation Learning for Source Code} 
Prior work has conducted many investigations to effectively represent the semantics of source code. Early studies widely use machine learning and traditional information retrieval methods. For instance, In \citep{DBLP:journals/ese/VasquezMPG14}, V\'{a}squez et al. adopt SVM to discriminate semantic similarities between code snippets and properly categorize software repositories. In the work \citep{DBLP:journals/tse/KamiyaKI02}, programs are morphed into token sequences for facilitating potential code clone detection. Recent work employs deep learning techniques for code semantics learning. Mou et al. adopt tree-structured convolutional neural network (Tree-CNN) to convert source code into distributed vector for program classiﬁcation \cite{DBLP:conf/aaai/MouLZWJ16}. The work \citep{DBLP:conf/msr/AkbarK19} suggests that the order of the embedded words can affect the accuracy of semantic representations. 
Besides the plain textual information, many studies utilize the structural features of source code, such as abstract syntax tree (AST) and control flow graph (CFG), to enrich the representations of source code. For example, in the work \citep{DBLP:conf/icse/ZhangWZ0WL19}, AST-based neural network is proposed to capture the structural information of source code. In another work~\cite{DBLP:conf/sigsoft/ChanCL12}, an API graph and a greedy subgraph search algorithm are utilized to help find the usage of source code, which excavates more semantic details in source code. Functions repetitively called and variables with the same names are involved into Graph Neural Networks (GNNs) for better representing the graph information in the work~\citep{DBLP:conf/iclr/AllamanisBK18}. Wan et al. propose a multi-modal attention network to combine the heterogeneous sources including AST, CFG and sequences of code tokens~\citep{DBLP:conf/kbse/WanSSXZ0Y19}.

\section{Conclusions}\label{sec:conclusion}
In this paper, we propose a novel deep neural network named \tool for code retrieval. According to our knowledge, \tool is the first deep learning model which utilizes the program dependency information for the task. \tool learns the code representations with the semantic dependency information combined. Specifically, the dependency information and statement-level tokens are jointly embedded for learning code semantics. Finally, \tool learns unified representations for both code and natural language queries. The experiment results have shown that \tool outperforms the state-of-the-art approaches and the semantic dependency learning is helpful for effective code retrieval. 

In the future, we will make a further exploration of the code structure and explicitly incorporate external knowledge such as API documentation to find a better way of representing source code semantics.

\bibliographystyle{cas-model2-names}

\bibliography{cas-refs}


\end{document}